\documentclass[10pt,twoside,twocolumn]{IEEEtran}

\IEEEoverridecommandlockouts
\usepackage{cite}
\usepackage{amsmath,amssymb,amsfonts}
\usepackage{algorithmic}
\usepackage{graphicx}
\usepackage{textcomp}
\usepackage{balance}
\usepackage{xcolor}
\def\BibTeX{{\rm B\kern-.05em{\sc i\kern-.025em b}\kern-.08em
    T\kern-.1667em\lower.7ex\hbox{E}\kern-.125emX}}

\usepackage{tikz}
\usepackage[utf8]{inputenc}
\usepackage{pgfplots}
\pgfplotsset{compat=newest}
\usepgfplotslibrary{groupplots}
\usepgfplotslibrary{dateplot}
\usepackage[caption=false,font=footnotesize]{subfig}
\usepackage{adjustbox}
\usepackage{amsmath}
\usepackage{cite}
\usepackage{microtype} 

\usepackage{times}

\newcommand{\eg}[1]{\textcolor{blue}{\bf [eg: #1]}}

\tikzset{every picture/.append style={line join=round,line cap=round,}}
\tikzstyle{solid}=[dash pattern=]
\tikzstyle{dotted}=[dash pattern=on 0.0\pgflinewidth off 2.5\pgflinewidth]
\tikzstyle{dashed}=[dash pattern=on 4.0\pgflinewidth off 3.0\pgflinewidth]
\tikzstyle{dashdotted}=[dash pattern=on 0.0\pgflinewidth off 2.0\pgflinewidth on 3.0\pgflinewidth off 2.0\pgflinewidth]
\pgfplotsset{every tick label/.append style={font=\large}}
\pgfplotsset{every axis legend/.append style={font=\Large}}
\pgfplotsset{every axis label/.append style={font=\Large}}

\usepackage{tikz}
\usetikzlibrary{shapes.geometric, arrows}
\usetikzlibrary{ arrows.meta, calc, positioning, quotes,}
\usepackage{circuitikz}
\usetikzlibrary{arrows.meta,positioning}
\usepackage{adjustbox}
\usepackage{todonotes}
\usepackage{soul}

\definecolor{bl1}{HTML}{F7FCF0}
\definecolor{bl2}{HTML}{E0F3DB}
\definecolor{bl3}{HTML}{CCEBC5}
\definecolor{bl4}{HTML}{A8DDB5}
\definecolor{bl5}{HTML}{7BCCC4}

\IEEEoverridecommandlockouts

\tikzstyle{startstop} = [rectangle, rounded corners, minimum width=0.5cm, minimum height=1cm,text centered, draw=black, fill=red!30]
\tikzstyle{io} = [trapezium, trapezium left angle=70, trapezium right angle=110, minimum width=0.5cm, minimum height=1cm, text centered, draw=black, fill=blue!30]
\tikzstyle{process} = [rectangle, minimum width=0.5cm, minimum height=1cm, text centered, text width=3cm, draw=black, fill=orange!30]
\tikzstyle{decision} = [diamond, minimum width=0.5cm, minimum height=1cm, text centered, draw=black, fill=green!30]
\tikzstyle{arrow} = [very thick,->,>=latex]
\tikzstyle{invisible} =[rectangle, node distance=3cm, text=white]

\usetikzlibrary{external}
\newcommand\bs{\mathbf{s}}

\newcommand\ba{\mathbf{a}}

\usepackage{amssymb}
\usepackage{amsfonts}
\usepackage{mathrsfs}
\usepackage{xspace}
\usepackage{bm}
\usepackage{upgreek}

\newcommand{\safemath}[2]{\newcommand{#1}{\ensuremath{#2}\xspace}}

\safemath{\bma}{\mathbf{a}}
\safemath{\bmb}{\mathbf{b}}
\safemath{\bmc}{\mathbf{c}}
\safemath{\bmd}{\mathbf{d}}
\safemath{\bme}{\mathbf{e}}
\safemath{\bmf}{\mathbf{f}}
\safemath{\bmg}{\mathbf{g}}
\safemath{\bmh}{\mathbf{h}}
\safemath{\bmi}{\mathbf{i}}
\safemath{\bmj}{\mathbf{j}}
\safemath{\bmk}{\mathbf{k}}
\safemath{\bml}{\mathbf{l}}
\safemath{\bmm}{\mathbf{m}}
\safemath{\bmn}{\mathbf{n}}
\safemath{\bmo}{\mathbf{o}}
\safemath{\bmp}{\mathbf{p}}
\safemath{\bmq}{\mathbf{q}}
\safemath{\bmr}{\mathbf{r}}
\safemath{\bms}{\mathbf{s}}
\safemath{\bmt}{\mathbf{t}}
\safemath{\bmu}{\mathbf{u}}
\safemath{\bmv}{\mathbf{v}}
\safemath{\bmw}{\mathbf{w}}
\safemath{\bmx}{\mathbf{x}}
\safemath{\bmy}{\mathbf{y}}
\safemath{\bmz}{\mathbf{z}}
\safemath{\bmzero}{\mathbf{0}}
\safemath{\bmone}{\mathbf{1}}

\bmdefine{\biad}{a}
\bmdefine{\bibd}{b}
\bmdefine{\bicd}{c}
\bmdefine{\bidd}{d}
\bmdefine{\bied}{e}
\bmdefine{\bifd}{f}
\bmdefine{\bigd}{g}
\bmdefine{\bihd}{h}
\bmdefine{\biid}{i}
\bmdefine{\bijd}{j}
\bmdefine{\bikd}{k}
\bmdefine{\bild}{l}
\bmdefine{\bimd}{m}
\bmdefine{\bind}{n}
\bmdefine{\biod}{o}
\bmdefine{\bipd}{p}
\bmdefine{\biqd}{q}
\bmdefine{\bird}{r}
\bmdefine{\bisd}{s}
\bmdefine{\bitd}{t}
\bmdefine{\biud}{u}
\bmdefine{\bivd}{v}
\bmdefine{\biwd}{w}
\bmdefine{\bixd}{x}
\bmdefine{\biyd}{y}
\bmdefine{\bizd}{z}

\bmdefine{\bixid}{\xi}
\bmdefine{\bilambdad}{\lambda}
\bmdefine{\bimud}{\mu}
\bmdefine{\bithetad}{\theta}
\bmdefine{\biphid}{\phi}
\bmdefine{\bideltad}{\delta}

\safemath{\bmia}{\biad}
\safemath{\bmib}{\bibd}
\safemath{\bmic}{\bicd}
\safemath{\bmid}{\bidd}
\safemath{\bmie}{\bied}
\safemath{\bmif}{\bifd}
\safemath{\bmig}{\bigd}
\safemath{\bmih}{\bihd}
\safemath{\bmii}{\biid}
\safemath{\bmij}{\bijd}
\safemath{\bmik}{\bikd}
\safemath{\bmil}{\bild}
\safemath{\bmim}{\bimd}
\safemath{\bmin}{\bind}
\safemath{\bmio}{\biod}
\safemath{\bmip}{\bipd}
\safemath{\bmiq}{\biqd}
\safemath{\bmir}{\bird}
\safemath{\bmis}{\bisd}
\safemath{\bmit}{\bitd}
\safemath{\bmiu}{\biud}
\safemath{\bmiv}{\bivd}
\safemath{\bmiw}{\biwd}
\safemath{\bmix}{\bixd}
\safemath{\bmiy}{\biyd}
\safemath{\bmiz}{\bizd}

\safemath{\bmxi}{\bixid}
\safemath{\bmlambda}{\bilambdad}
\safemath{\bmmu}{\bimud}
\safemath{\bmtheta}{\bithetad}
\safemath{\bmphi}{\biphid}
\safemath{\bmdelta}{\bideltad}

\safemath{\bA}{\mathbf{A}}
\safemath{\bB}{\mathbf{B}}
\safemath{\bC}{\mathbf{C}}
\safemath{\bD}{\mathbf{D}}
\safemath{\bE}{\mathbf{E}}
\safemath{\bF}{\mathbf{F}}
\safemath{\bG}{\mathbf{G}}
\safemath{\bH}{\mathbf{H}}
\safemath{\bI}{\mathbf{I}}
\safemath{\bJ}{\mathbf{J}}
\safemath{\bK}{\mathbf{K}}
\safemath{\bL}{\mathbf{L}}
\safemath{\bM}{\mathbf{M}}
\safemath{\bN}{\mathbf{N}}
\safemath{\bO}{\mathbf{O}}
\safemath{\bP}{\mathbf{P}}
\safemath{\bQ}{\mathbf{Q}}
\safemath{\bR}{\mathbf{R}}
\safemath{\bS}{\mathbf{S}}
\safemath{\bT}{\mathbf{T}}
\safemath{\bU}{\mathbf{U}}
\safemath{\bV}{\mathbf{V}}
\safemath{\bW}{\mathbf{W}}
\safemath{\bX}{\mathbf{X}}
\safemath{\bY}{\mathbf{Y}}
\safemath{\bZ}{\mathbf{Z}}

\safemath{\bZero}{\mathbf{0}}
\safemath{\bOne}{\mathbf{1}}
\safemath{\bDelta}{\mathbf{\Delta}}
\safemath{\bLambda}{\mathbf{\UpLambda}}
\safemath{\bPhi}{\mathbf{\Phi}}
\safemath{\bPsi}{\mathbf{\Psi}}
\safemath{\bSigma}{\mathbf{\Upsigma}}
\safemath{\bOmega}{\mathbf{\Upomega}}
\safemath{\bTheta}{\mathbf{\Uptheta}}

\bmdefine{\biAd}{A}
\bmdefine{\biBd}{B}
\bmdefine{\biCd}{C}
\bmdefine{\biDd}{D}
\bmdefine{\biEd}{E}
\bmdefine{\biFd}{F}
\bmdefine{\biGd}{G}
\bmdefine{\biHd}{H}
\bmdefine{\biId}{I}
\bmdefine{\biJd}{J}
\bmdefine{\biKd}{K}
\bmdefine{\biLd}{L}
\bmdefine{\biMd}{M}
\bmdefine{\biOd}{N}
\bmdefine{\biPd}{O}
\bmdefine{\biQd}{P}
\bmdefine{\biRd}{R}
\bmdefine{\biSd}{S}
\bmdefine{\biTd}{T}
\bmdefine{\biUd}{U}
\bmdefine{\biVd}{V}
\bmdefine{\biWd}{W}
\bmdefine{\biXd}{X}
\bmdefine{\biYd}{Y}
\bmdefine{\biZd}{Z}

\bmdefine{\biDelta}{\Delta}
\bmdefine{\biLambda}{\Lambda}
\bmdefine{\biPhi}{\Phi}
\bmdefine{\biSigma}{\Sigma}
\bmdefine{\biOmega}{\Omega}
\bmdefine{\biTheta}{\Theta}

\safemath{\bimA}{\biAd}
\safemath{\bimB}{\biBd}
\safemath{\bimC}{\biCd}
\safemath{\bimD}{\biDd}
\safemath{\bimE}{\biEd}
\safemath{\bimF}{\biFd}
\safemath{\bimG}{\biGd}
\safemath{\bimH}{\biHd}
\safemath{\bimI}{\biId}
\safemath{\bimJ}{\biJd}
\safemath{\bimK}{\biKd}
\safemath{\bimL}{\biLd}
\safemath{\bimM}{\biMd}
\safemath{\bimN}{\biNd}
\safemath{\bimO}{\biOd}
\safemath{\bimP}{\biPd}
\safemath{\bimQ}{\biQd}
\safemath{\bimR}{\biRd}
\safemath{\bimS}{\biSd}
\safemath{\bimT}{\biTd}
\safemath{\bimU}{\biUd}
\safemath{\bimV}{\biVd}
\safemath{\bimW}{\biWd}
\safemath{\bimX}{\biXd}
\safemath{\bimY}{\biYd}
\safemath{\bimZ}{\biZd}

\safemath{\bimDelta}{\biDelta}
\safemath{\bimLambda}{\biLambda}
\safemath{\bimPhi}{\biPhi}
\safemath{\bimSigma}{\biSigma}
\safemath{\bimOmega}{\biOmega}
\safemath{\bimTheta}{\biTheta}

\safemath{\setA}{\mathcal{A}}
\safemath{\setB}{\mathcal{B}}
\safemath{\setC}{\mathcal{C}}
\safemath{\setD}{\mathcal{D}}
\safemath{\setE}{\mathcal{E}}
\safemath{\setF}{\mathcal{F}}
\safemath{\setG}{\mathcal{G}}
\safemath{\setH}{\mathcal{H}}
\safemath{\setI}{\mathcal{I}}
\safemath{\setJ}{\mathcal{J}}
\safemath{\setK}{\mathcal{K}}
\safemath{\setL}{\mathcal{L}}
\safemath{\setM}{\mathcal{M}}
\safemath{\setN}{\mathcal{N}}
\safemath{\setO}{\mathcal{O}}
\safemath{\setP}{\mathcal{P}}
\safemath{\setQ}{\mathcal{Q}}
\safemath{\setR}{\mathcal{R}}
\safemath{\setS}{\mathcal{S}}
\safemath{\setT}{\mathcal{T}}
\safemath{\setU}{\mathcal{U}}
\safemath{\setV}{\mathcal{V}}
\safemath{\setW}{\mathcal{W}}
\safemath{\setX}{\mathcal{X}}
\safemath{\setY}{\mathcal{Y}}
\safemath{\setZ}{\mathcal{Z}}
\safemath{\emptySet}{\varnothing}

\safemath{\colA}{\mathscr{A}}
\safemath{\colB}{\mathscr{B}}
\safemath{\colC}{\mathscr{C}}
\safemath{\colD}{\mathscr{D}}
\safemath{\colE}{\mathscr{E}}
\safemath{\colF}{\mathscr{F}}
\safemath{\colG}{\mathscr{G}}
\safemath{\colH}{\mathscr{H}}
\safemath{\colI}{\mathscr{I}}
\safemath{\colJ}{\mathscr{J}}
\safemath{\colK}{\mathscr{K}}
\safemath{\colL}{\mathscr{L}}
\safemath{\colM}{\mathscr{M}}
\safemath{\colN}{\mathscr{N}}
\safemath{\colO}{\mathscr{O}}
\safemath{\colP}{\mathscr{P}}
\safemath{\colQ}{\mathscr{Q}}
\safemath{\colR}{\mathscr{R}}
\safemath{\colS}{\mathscr{S}}
\safemath{\colT}{\mathscr{T}}
\safemath{\colU}{\mathscr{U}}
\safemath{\colV}{\mathscr{V}}
\safemath{\colW}{\mathscr{W}}
\safemath{\colX}{\mathscr{X}}
\safemath{\colY}{\mathscr{Y}}
\safemath{\colZ}{\mathscr{Z}}

\safemath{\opA}{\mathbb{A}}
\safemath{\opB}{\mathbb{B}}
\safemath{\opC}{\mathbb{C}}
\safemath{\opD}{\mathbb{D}}
\safemath{\opE}{\mathbb{E}}
\safemath{\opF}{\mathbb{F}}
\safemath{\opG}{\mathbb{G}}
\safemath{\opH}{\mathbb{H}}
\safemath{\opI}{\mathbb{I}}
\safemath{\opJ}{\mathbb{J}}
\safemath{\opK}{\mathbb{K}}
\safemath{\opL}{\mathbb{L}}
\safemath{\opM}{\mathbb{M}}
\safemath{\opN}{\mathbb{N}}
\safemath{\opO}{\mathbb{O}}
\safemath{\opP}{\mathbb{P}}
\safemath{\opQ}{\mathbb{Q}}
\safemath{\opR}{\mathbb{R}}
\safemath{\opS}{\mathbb{S}}
\safemath{\opT}{\mathbb{T}}
\safemath{\opU}{\mathbb{U}}
\safemath{\opV}{\mathbb{V}}
\safemath{\opW}{\mathbb{W}}
\safemath{\opX}{\mathbb{X}}
\safemath{\opY}{\mathbb{Y}}
\safemath{\opZ}{\mathbb{Z}}
\safemath{\opZero}{\mathbb{O}}
\safemath{\identityop}{\opI}

\safemath{\veca}{\bma}
\safemath{\vecb}{\bmb}
\safemath{\vecc}{\bmc}
\safemath{\vecd}{\bmd}
\safemath{\vece}{\bme}
\safemath{\vecf}{\bmf}
\safemath{\vecg}{\bmg}
\safemath{\vech}{\bmh}
\safemath{\veci}{\bmi}
\safemath{\vecj}{\bmj}
\safemath{\veck}{\bmk}
\safemath{\vecl}{\bml}
\safemath{\vecm}{\bmm}
\safemath{\vecn}{\bmn}
\safemath{\veco}{\bmo}
\safemath{\vecp}{\bmp}
\safemath{\vecq}{\bmq}
\safemath{\vecr}{\bmr}
\safemath{\vecs}{\bms}
\safemath{\vect}{\bmt}
\safemath{\vecu}{\bmu}
\safemath{\vecv}{\bmv}
\safemath{\vecw}{\bmw}
\safemath{\vecx}{\bmx}
\safemath{\vecy}{\bmy}
\safemath{\vecz}{\bmz}

\safemath{\veczero}{\bmzero}
\safemath{\vecone}{\bmone}
\safemath{\vecxi}{\bmxi}
\safemath{\veclambda}{\bmlambda}
\safemath{\vecmu}{\bmmu}
\safemath{\vectheta}{\bmtheta}
\safemath{\vecphi}{\bmphi}
\safemath{\vecdelta}{\bmdelta}

\safemath{\matA}{\bA}
\safemath{\matB}{\bB}
\safemath{\matC}{\bC}
\safemath{\matD}{\bD}
\safemath{\matE}{\bE}
\safemath{\matF}{\bF}
\safemath{\matG}{\bG}
\safemath{\matH}{\bH}
\safemath{\matI}{\bI}
\safemath{\matJ}{\bJ}
\safemath{\matK}{\bK}
\safemath{\matL}{\bL}
\safemath{\matM}{\bM}
\safemath{\matN}{\bN}
\safemath{\matO}{\bO}
\safemath{\matP}{\bP}
\safemath{\matQ}{\bQ}
\safemath{\matR}{\bR}
\safemath{\matS}{\bS}
\safemath{\matT}{\bT}
\safemath{\matU}{\bU}
\safemath{\matV}{\bV}
\safemath{\matW}{\bW}
\safemath{\matX}{\bX}
\safemath{\matY}{\bY}
\safemath{\matZ}{\bZ}
\safemath{\matzero}{\bmzero}

\safemath{\matDelta}{\bDelta}
\safemath{\matLambda}{\bLambda}
\safemath{\matPhi}{\bPhi}
\safemath{\matSigma}{\bSigma}
\safemath{\matOmega}{\bOmega}
\safemath{\matTheta}{\bTheta}

\safemath{\matidentity}{\matI}
\safemath{\matone}{\matO}

\safemath{\rnda}{A}
\safemath{\rndb}{B}
\safemath{\rndc}{C}
\safemath{\rndd}{D}
\safemath{\rnde}{E}
\safemath{\rndf}{F}
\safemath{\rndg}{G}
\safemath{\rndh}{H}
\safemath{\rndi}{I}
\safemath{\rndj}{J}
\safemath{\rndk}{K}
\safemath{\rndl}{L}
\safemath{\rndm}{M}
\safemath{\rndn}{N}
\safemath{\rndo}{O}
\safemath{\rndp}{P}
\safemath{\rndq}{Q}
\safemath{\rndr}{R}
\safemath{\rnds}{S}
\safemath{\rndt}{T}
\safemath{\rndu}{U}
\safemath{\rndv}{V}
\safemath{\rndw}{W}
\safemath{\rndx}{X}
\safemath{\rndy}{Y}
\safemath{\rndz}{Z}

\safemath{\rveca}{\bimA}
\safemath{\rvecb}{\bimB}
\safemath{\rvecc}{\bimC}
\safemath{\rvecd}{\bimD}
\safemath{\rvece}{\bimE}
\safemath{\rvecf}{\bimF}
\safemath{\rvecg}{\bimG}
\safemath{\rvech}{\bimH}
\safemath{\rveci}{\bimI}
\safemath{\rvecj}{\bimJ}
\safemath{\rveck}{\bimK}
\safemath{\rvecl}{\bimL}
\safemath{\rvecm}{\bimM}
\safemath{\rvecn}{\bimN}
\safemath{\rveco}{\bomO}
\safemath{\rvecp}{\bimP}
\safemath{\rvecq}{\bimQ}
\safemath{\rvecr}{\bimR}
\safemath{\rvecs}{\bimS}
\safemath{\rvect}{\bimT}
\safemath{\rvecu}{\bimU}
\safemath{\rvecv}{\bimV}
\safemath{\rvecw}{\bimW}
\safemath{\rvecx}{\bimX}
\safemath{\rvecy}{\bimY}
\safemath{\rvecz}{\bimZ}

\safemath{\rvecxi}{\bmxi}
\safemath{\rveclambda}{\bmlambda}
\safemath{\rvecmu}{\bmmu}
\safemath{\rvectheta}{\bmtheta}
\safemath{\rvecphi}{\bmphi}

\safemath{\rmatA}{\bimA}
\safemath{\rmatB}{\bimB}
\safemath{\rmatC}{\bimC}
\safemath{\rmatD}{\bimD}
\safemath{\rmatE}{\bimE}
\safemath{\rmatF}{\bimF}
\safemath{\rmatG}{\bimG}
\safemath{\rmatH}{\bimH}
\safemath{\rmatI}{\bimI}
\safemath{\rmatJ}{\bimJ}
\safemath{\rmatK}{\bimK}
\safemath{\rmatL}{\bimL}
\safemath{\rmatM}{\bimM}
\safemath{\rmatN}{\bimN}
\safemath{\rmatO}{\bimO}
\safemath{\rmatP}{\bimP}
\safemath{\rmatQ}{\bimQ}
\safemath{\rmatR}{\bimR}
\safemath{\rmatS}{\bimS}
\safemath{\rmatT}{\bimT}
\safemath{\rmatU}{\bimU}
\safemath{\rmatV}{\bimV}
\safemath{\rmatW}{\bimW}
\safemath{\rmatX}{\bimX}
\safemath{\rmatY}{\bimY}
\safemath{\rmatZ}{\bimZ}

\safemath{\rmatDelta}{\bimDelta}
\safemath{\rmatLambda}{\bimLambda}
\safemath{\rmatPhi}{\bimPhi}
\safemath{\rmatSigma}{\bimSigma}
\safemath{\rmatOmega}{\bimOmega}
\safemath{\rmatTheta}{\bimTheta}

\usepackage{amssymb}
\usepackage{amsfonts}
\usepackage{mathrsfs}
\usepackage{xspace}
\usepackage{bm}
\usepackage{fancyref}
\usepackage{textcomp}

\usepackage{multirow}
\usepackage{stmaryrd}

\newenvironment{textbmatrix}{	\setlength{\arraycolsep}{2.5pt}%
								\big[\begin{matrix}}{\end{matrix}\big]%
								\raisebox{0.08ex}{\vphantom{M}}}

\def\be{\begin{equation}}
\def\ee{\end{equation}}
\def\een{\nonumber \end{equation}}
\def\mat{\begin{bmatrix}}
\def\emat{\end{bmatrix}}
\def\btm{\begin{textbmatrix}}
\def\etm{\end{textbmatrix}}

\def\ba#1\ea{\begin{align}#1\end{align}}
\def\bas#1\eas{\begin{align*}#1\end{align*}}
\def\bs#1\es{\begin{split}#1\end{split}} 
\def\bg#1\eg{\begin{gather}#1\end{gather}}
\def\bml#1\eml{\begin{multline}#1\end{multline}}
\def\bi#1\ei{\begin{itemize}#1\end{itemize}}

\DeclareMathOperator*{\argmax}{arg\;max}		

\safemath{\dirac}{\delta}					
\safemath{\krond}{\dirac}					

\safemath{\upto}{\uparrow}
\safemath{\downto}{\downarrow}
\safemath{\iu}{j}							
\safemath{\ev}{\lambda}						
\safemath{\hilseqspace}{l^{2}}				
\newcommand{\banachfunspace}[1]{\setL^{#1}}	
\safemath{\hilfunspace}{\banachfunspace{2}}	

\safemath{\SNR}{\text{\sc snr}} 				
\safemath{\No}{N_0}							
\safemath{\Es}{E_s}							
\safemath{\Eb}{E_b}							
\safemath{\EbNo}{\frac{\Eb}{\No}}
\safemath{\EsNo}{\frac{\Es}{\No}}

\DeclareMathOperator{\CHop}{\ensuremath{\opH}} 
\safemath{\tvir}{\rndh_{\CHop}}				
\safemath{\tvtf}{\rndl_{\CHop}}				
\safemath{\spf}{\rnds_{\CHop}}				
\safemath{\bff}{H_{\CHop}}					

\safemath{\ircf}{r_{h}}						
\safemath{\tftvcf}{r_{s}}					
\safemath{\tfcf}{r_{l}}						
\safemath{\bfcf}{r_{H}}						

\safemath{\tcorr}{c_h}						
\safemath{\scf}{c_{s}}						
\safemath{\tfcorr}{c_{l}}					
\safemath{\fcorr}{c_{H}}						

\safemath{\mi}{I}							
\safemath{\capacity}{C}						

\safemath{\normal}{\mathcal{N}}			
\safemath{\jpg}{\mathcal{CN}}			
\safemath{\mchain}{\leftrightarrow}		

\safemath{\dB}{\,\mathrm{dB}}
\safemath{\dBm}{\,\mathrm{dBm}}
\safemath{\Hz}{\,\mathrm{Hz}}
\safemath{\kHz}{\,\mathrm{kHz}}
\safemath{\MHz}{\,\mathrm{MHz}}
\safemath{\GHz}{\,\mathrm{GHz}}
\safemath{\s}{\,\mathrm{s}}
\safemath{\ms}{\,\mathrm{ms}}
\safemath{\mus}{\,\mathrm{\text{\textmu}s}}
\safemath{\ns}{\,\mathrm{ns}}
\safemath{\ps}{\,\mathrm{ps}}
\safemath{\meter}{\,\mathrm{m}}
\safemath{\mm}{\,\mathrm{mm}}
\safemath{\cm}{\,\mathrm{cm}}
\safemath{\W}{\,\mathrm{W}}
\safemath{\mW}{\, \mathrm{mW}}
\safemath{\J}{\,\mathrm{J}}
\safemath{\K}{\,\mathrm{K}}
\safemath{\bit}{\,\mathrm{bit}}
\safemath{\nat}{\,\mathrm{nat}}

\safemath{\define}{\triangleq}			

\safemath{\equivalent}{\sim}
\safemath{\distas}{\sim}					
\safemath{\sdiff}{\Delta}				

\safemath{\reals}{\mathbb{R}}
\safemath{\positivereals}{\reals_{+}}
\safemath{\integers}{\mathbb{Z}}
\safemath{\posint}{\integers_{+}}
\safemath{\naturals}{\mathbb{N}}
\safemath{\posnaturals}{\naturals_{+}}
\safemath{\complexset}{\mathbb{C}}
\safemath{\rationals}{\mathbb{Q}}

\newcommand*{\fancyrefapplabelprefix}{app}		
\newcommand*{\fancyrefthmlabelprefix}{thm}		
\newcommand*{\fancyreflemlabelprefix}{lem}		
\newcommand*{\fancyrefcorlabelprefix}{cor}		
\newcommand*{\fancyrefdeflabelprefix}{def}		
\newcommand*{\fancyrefproplabelprefix}{prop}		
\newcommand*{\fancyrefexmpllabelprefix}{exmpl}
\frefformat{vario}{\fancyrefseclabelprefix}{Section~#1}
\frefformat{vario}{\fancyrefthmlabelprefix}{Theorem~#1}
\frefformat{vario}{\fancyreflemlabelprefix}{Lemma~#1}
\frefformat{vario}{\fancyrefcorlabelprefix}{Corollary~#1}
\frefformat{vario}{\fancyrefdeflabelprefix}{Definition~#1}
\frefformat{vario}{\fancyreffiglabelprefix}{Fig.~#1}
\frefformat{vario}{\fancyrefapplabelprefix}{Appendix~#1}
\frefformat{vario}{\fancyrefeqlabelprefix}{(#1)}
\frefformat{vario}{\fancyrefproplabelprefix}{Property~#1}
\frefformat{vario}{\fancyrefexmpllabelprefix}{Example~#1}

\begin{document}

\title{Attacking and Defending Deep-Learning-Based Off-Device Wireless Positioning Systems\\

}

\author{\IEEEauthorblockN{Pengzhi Huang, Emre G\"{o}n\"{u}lta\c{s}, Maximilian Arnold, K. Pavan Srinath, Jakob Hoydis, and Christoph Studer}%
\thanks{%
P.~Huang is with the School of Electrical and Computer Engineering, Cornell University (ph448@cornell.edu). 
E.~G\"{o}n\"{u}lta\c{s} was with the School of Electrical and Computer Engineering, Cornell University, Ithaca, NY. He is now with Ericsson, Austin,
TX (emre.gonultas@ericsson.com). 
M.~Arnold is with  Qualcomm AI-Research, Netherlands (marnold@qti.qualcomm.com).
K.~P.~Srinath is with Nokia Bell Labs, France (pavan.koteshwar\_srinath@nokia-bell-labs.com). 
J.~Hoydis is with NVIDIA Corp.\ (jhoydis@nvidia.com). 
C.~Studer is with ETH Zurich, Z\"urich, Switzerland (studer@ethz.ch).}%
\thanks{The work of CS was supported in part by ComSenTer, one of six centers in JUMP, a SRC program sponsored by DARPA,  by the U.S.\ NSF under grants CNS-1717559 and ECCS-1824379, by the Swiss NSF grant 207314, by the CHIST-ERA project CHASER through the Swiss NSF grant 218704, and by an ETH Zurich Research grant.}
}

\maketitle

\begin{abstract}
Localization services for wireless devices play an increasingly important role in our daily lives and a plethora of emerging services and applications already rely on precise position information. Widely used on-device positioning methods, such as the global positioning system, enable accurate outdoor positioning and provide the users with full control over what services and applications are allowed to access their location information. In order to provide accurate positioning indoors or in cluttered urban scenarios without line-of-sight satellite connectivity, powerful off-device  positioning systems, which process channel state information (CSI) measured at the infrastructure base stations or access points with deep neural networks, have emerged recently. Such off-device wireless positioning systems inherently link a user's data transmission with its localization, since accurate CSI measurements are necessary for reliable wireless communication---this not only prevents the users from controlling who can access this information but also enables virtually everyone in the device's range to estimate its location, resulting in serious privacy and security concerns. We therefore propose on-device attacks against off-device wireless positioning systems in multi-antenna orthogonal frequency-division multiplexing systems while remaining standard compliant and minimizing the impact on quality-of-service, and we demonstrate their efficacy using real-world measured datasets for cellular outdoor and wireless LAN indoor scenarios. We also investigate defenses to counter such attack mechanisms, and we  discuss the limitations and implications on protecting location privacy in existing and future wireless communication systems. 
\end{abstract}


\section{Introduction}
\IEEEPARstart{D}{evice} localization services play an important role in society as they are widely used in applications ranging from car navigation, augmented reality, targeted advertisement, asset tracking, health and lifestyle apps, industry automation, and many more~\cite{han2016survey,wicker2012locationprivacy}.
Existing positioning systems can be categorized as follows: (i) {on-device} systems that enable the users to locate themselves and (ii) {off-device} systems that enable the receiver to locate the users. 
On-device positioning based on global navigation satellite systems (GNSSs), such as the global positioning system~\cite{hofmann2012global}, provides accurate location estimates, and privacy mechanisms on modern mobile devices typically provide full control over what services are allowed to gain access to the acquired GNSS coordinates (e.g., a car navigation app)~\cite{orman2013privacy}.
Such GNSS-based systems are the de-facto standard for outdoor localization but require line-of-sight connectivity to multiple satellites for proper functioning, rendering them ineffective for indoor applications.
As a remedy, off-device alternatives that process communication signals have been proposed for indoor localization~\cite{WEN201921survey,gustaffson2005wirelesspos,sahinoglu2008ultra,studer20205g}.
Emerging from the network provider, indoor positioning techniques that rely on wave-propagation or geometric models to process time-of-flight (ToF) or angle-of-arrival (AoA) measurements at base-stations (BSs) or access points (APs) with multiple antennas have been proposed for regulatory (e.g., E911) and commercial services~\cite{SurveyLocalizationTechniques}.  
The recent trend towards more antennas and increased bandwidth led to the definition of performance requirements for fifth generation (5G) wireless standards~\cite{3gpp5grel16position}, allowing the design of localization systems that exploit the existing communication link.
Moreover, future standard releases are expected to specify even more stringent specifications in positioning accuracy~\cite{larsson2015fingerprinting,keating20195gpositioning}.

Besides such off-device localization methods that rely on wave-propagation models, powerful approaches that process wireless signals using deep neural networks (DNNs) have emerged recently~\cite{wang2015deepfi,chen2017confi,wang2017csi,lundpaper,arnold2019novel,zappone2019surverwireless,wang2020bimodalcsi,huawei2020paper,gonultacs2021csi,foliadis2021csibased,li2021indoor}.
Such positioning systems not only provide accurate indoor and outdoor position estimates where ToF and AoA measurements are unreliable (e.g., in non-line-of-sight or high-mobility scenarios), but also enable infrastructure BSs or APs to localize wireless transmitters without the users' consent. Thus, network operators are able to locate any device that connects to their infrastructure. 
Furthermore, as wireless signals can be measured unnoticed by virtually everyone in the transmitter's range, it is possible to design malicious systems that passively sniff wireless signals to perform off-device localization, without even being part of the network~\cite{rpicsi2019}.  
Consequently, DNN-based wireless positioning causes a number of serious privacy and security concerns~\cite{han2016survey,wicker2012locationprivacy}, which calls for effective attack mechanisms that prevent accurate device localization while remaining compliant to the communication standard and maintaining high quality-of-service in terms of data rates, coverage, and range. We address exactly these aspects in this paper.

\subsection{The Basics of CSI-Based Wireless Positioning}
In order to enable reliable data transmission, wireless systems typically estimate the channel's transfer function (e.g., in the form of impulse responses) using a training sequence that is known to both the transmitter and receiver. This acquired channel state information (CSI) is then used to inverse the channel's effect with the goals of detecting the transmitted data.
Clearly, the measured CSI contains information about the propagation environment, which can be utilized for wireless positioning~\cite{zappone2019surverwireless,lundpaper,huawei2020paper,gonultacs2021csi}; this is an immediate consequence of the fact that wireless signals are affected by the physical environment between the transmitter and receiver, including attenuation and ToF over distance, reflections through scattering, and/or refraction through certain materials. 
While CSI must be acquired continuously in order to ensure reliable communication, it perpetually provides location information of the transmitting device.

Traditional wireless positioning techniques utilize geometrical models for wave propagation in order to perform triangulation or trilateration~\cite{rosado2018mobilepositioningsurvey}. Such methods rely on either AoA (e.g., using an array of antennas) and/or ToF information (e.g., using precise BS/AP synchronization) to localize the transmitting device. 
Unfortunately, such approaches require  antenna calibration and accurate time synchronization between the receivers, and require mostly idealistic line-of-sight propagation conditions between transmitter and receiver, which often prevents accurate positioning indoors or in dense urban scenarios that contain multiple scatterers.

More recently, model-free wireless positioning systems that process the measured CSI with the aid of DNNs have emerged~\cite{wang2015deepfi,wang2017csi,chen2017confi,lundpaper,ericpaper,arnold2019novel,huawei2020paper,gonultacs2021csi,gonultacs2021feature,li2021wireless,foliadis2021csibased,li2021indoor}. 
Their principle is simple yet effective: Train a DNN with a dataset consisting of CSI measurements and ground-truth locations obtained via a reference positioning system in the area of interest. Then, use the trained DNN to process new CSI measurements to generate location estimates.
Such positioning systems enable accurate localization even under challenging propagation conditions that prevent approaches relying on geometrical models. Experiments with real-world data have demonstrated meter-level and centimeter-level accuracy for outdoor~\cite{huawei2020paper} and indoor~\cite{gonultacs2021csi,gonultacs2021feature,foliadis2021csibased} scenarios, respectively.

\subsection{Privacy and Security Threats}
Off-device wireless positioning is a blessing and a curse.
Since the extraction of location estimates can be carried out on server farms, off-device positioning methods can utilize computationally complex algorithms and save energy on battery-powered wireless devices. However, modern communication standards~\cite{ieee80211ac,3gpp5grel15}   enable virtually everyone in the transmitters' vicinity to acquire CSI and, hence, extract location estimates. 
Furthermore, off-device positioning systems inherently link a user’s data transmission with the acquisition of CSI, which is necessary for estimating the transmitted data; this implies that preventing the acquisition of CSI at the receiver side is typically equivalent to preventing communication. 

By extracting location estimates from CSI, privacy is at stake. For example, one can learn the users' behavior (e.g., what stores or bars they frequently visit), their socioeconomic status (e.g., based on where they live and work), 
or even gain sensitive information about their health (e.g., if the user frequently visits a specialized health clinic)~\cite{wicker2012locationprivacy}.
One can even envision concrete security threats, as location information enables one to determine when a user is leaving their home, parking their car, or withdrawing {cash}. 
It is therefore crucial to develop effective countermeasures that prevent accurate off-device localization without sacrificing quality-of-service.

\subsection{Contributions}
We investigate on-device attacks in order to avoid accurate CSI-based localization with DNNs, while remaining standard compliant and enabling reliable communication.
Our main contributions are as follows:
\begin{itemize}
\item We develop a simple, yet effective user-side attack to off-device CSI-based positioning systems that convolves the transmit signals with short perturbation sequences in order to prevent accurate wireless localization in systems using orthogonal frequency division multiplexing (OFDM).
\item We propose a range of (deterministic) adversarial as well as randomized attacks with varying knowledge of the positioning pipeline and the wireless channel in order to generate suitable perturbation sequences that maximize the positioning error.  
\item We investigate defense mechanisms that render off-device CSI-based positioning systems more resilient to the proposed attacks.
\item We demonstrate the efficacy of our methods using two measured datasets in a 3GPP-5G release~15~\cite{3gpp5grel15} outdoor scenario and an IEEE 802.11ac WLAN~\cite{ieee80211ac} indoor scenario, and we explore the trade-offs between robustness, localization privacy, and communication capacity.
\item We conclude by discussing the limits and implications of our results on existing and emerging wireless communication systems.
\end{itemize}

\subsection{Relevant Prior Work}\label{sec:Prior}

Ever since adversarial attacks were discovered as potential vulnerabilities of DNN-based image classifiers~\cite{szegedy2014intriguing,biggio2013evasion}, a range of defensive mechanisms have been proposed, e.g.,~\cite{papernot2016distillation,raghunathan2018certified,samangouei2018defense}.
While powerful counterattacks keep on emerging, e.g.,~\cite{narodytska2016simple,chen2020hopskipjumpattack,madry2017towards},  
theory indicates that adversarial attacks are inevitable for general, high-dimensional classifiers~\cite{shafahi2018adversarial,ilyas2019adversarial} and, hence, inherently difficult to avoid.
In contrast to the challenges adversarial attacks pose to many classification tasks, our results for CSI-based wireless positioning indicate that their existence is an opportunity, as they can be exploited to improve location privacy.

Location privacy is not a novel topic; see, e.g.,~\cite{wicker2012locationprivacy,di2010privacywireless,WiFiProtect ,WifiResponse}. Existing work largely focuses on the trade-offs between location accuracy provided by the users (obtained from on-device positioning systems) versus the fidelity of the services provided to the user~\cite{danezis2005much,krumm2009survey}. 
{In contrast to such results, we focus on off-device positioning systems that process measured CSI in order to extract location estimates using DNNs.} Such systems circumvent most of these trade-offs as localization is possible without the users' consent~\cite{WEN201921survey,gustaffson2005wirelesspos,sahinoglu2008ultra,studer20205g}.

Privacy issues also arise with widely-available wireless sniffers that passively collect unencrypted physical layer information, including medium access control (MAC) addresses~\cite{rpicsi2019}.
In order to attack user identification, potential countermeasures, such as continuously changing the MAC address, have been proposed in \cite{Zhu_2020wifisensing}.
In contrast to such MAC-layer approaches, we focus on simple yet effective physical (PHY) layer techniques at the UE-side that indirectly manipulate the CSI estimated at the BS or AP in order to prevent accurate localization.

WiFi-based positioning systems that rely on geometrical models with ToF and AoA measurements where among the first ones for which countermeasures were developed \cite{WiFiProtect,qiao2016phycloak}. 
Shortly after, methods that mitigate such attacks have emerged \cite{WiFiProtect}, causing a race between measures and countermeasures~\cite{WifiResponse}. 
However, algorithmic methods that attack geometry-based ToF and AoA-based positioning systems are generally ineffective against DNN-based approaches as the learned function that performs positioning is described by a black box. 
Hence, we propose countermeasures that are specifically designed to attack DNN-based systems.

For DNN-based positioning systems, methods that manipulate the frequency-domain training symbols of orthogonal frequency-division multiplexing (OFDM) systems have been proposed in~\cite{abanto2020wifi,COMINELLI2021107970,cominelli2022properties}.
However, directly manipulating the frequency-domain information at the transmitter (before the inverse discrete Fourier transform) is likely to violate the mutual orthogonality of the OFDM {subcarriers} as the effective channel impulse response may exceed the cyclic prefix length---{this will negatively affect the quality of service}. 
{In contrast, we propose time-domain attacks that preconvolve the transmitted signals by a specific perturbation sequence of pre-defined (and short) length, which has a well-defined impact on the received CSI while maintaining subcarrier orthogonality as long as the effective channel's delay spread does not exceed the cyclic prefix length. Furthermore, we consider different threat models with varying knowledge of the DNN-based positioning pipeline, which trade off model knowledge versus efficacy.}
{Besides those differences, we also investigate defensive mechanisms that can be deployed at the infrastructure BSs/APs---this aspect is, to the best of our knowledge, novel.}

Reference  \cite{zhou2019adversarial} proposes adversarial attacks on the frequency-domain CSI against WiFi-based human activity detection. {This approach, however, relies on traditional DNN-based classifier models without considering compatibility with a communication standard.} 
In contrast, {our attacks are targeting positioning models (which typically perform inference instead of classification) and addressing compatibility with existing OFDM communication systems. Furthermore, we propose a range of adversarial attacks with varying knowledge on the positioning pipeline, which specifically target CSI-based localization using DNNs.}

We conclude by noting that our adversarial attacks resemble that of functional attacks~\cite{laidlaw2019functional}, where the adversary applies a carefully-designed function to each pixel of a picture, perturbing the same color by the same amount. Analogously, we convolve the transmit-side time-domain signals with well-designed perturbation sequences, which affects the CSI estimated at the receiver by an entry-wise multiplication with the channel's transfer function.

\subsection{Notation}

Lowercase boldface, uppercase boldface, and uppercase calligraphic letters denote column vectors, matrices, and sets, respectively. 
The $i$th element of a vector $\bma$ is denoted by $a_i$, the $i$th column of a matrix $\bA$ by $\bma_i$, and the element on the $i$th row and $j$th column by $A_{i,j}$. 
The $\ell^2$-norm of a vector $\bma$ is $\|\mathbf{a}\|$.
The transpose, Hermitian transpose, and Frobenius norm of a matrix $\bA$ is $\mathbf{A}^T$, $\mathbf{A}^H$, and $\|\mathbf{A}\|_F$, respectively. 
The Hadamard product between the matrices $\mathbf{A}$ and $\mathbf{B}$ is denoted by $\mathbf{A}\odot\mathbf{B}$. 
The convolution and circular convolution between the vectors~$\bma$ and~$\bmb$ is denoted by $\bma\ast\bmb$ and $\bma\circledast\bmb$, respectively.

\section{Communication System and CSI-Based Wireless Positioning}
\label{section:System Model}
In what follows, we focus on communication systems that rely on OFDM~\cite{ofdm,li2006orthogonal}, as they are widely used in modern wireless communication standards, such as 3GPP-LTE \cite{lte}, 3GPP-5G~\cite{3gpp5grel15}, and wireless LAN (WLAN)~\cite{ieee80211ac}. 
We first detail the system model and then, introduce the principles of off-device CSI-based positioning.

\begin{figure*}
    \centering
\adjustbox{max width=0.99\textwidth,valign=t}{%
\includegraphics{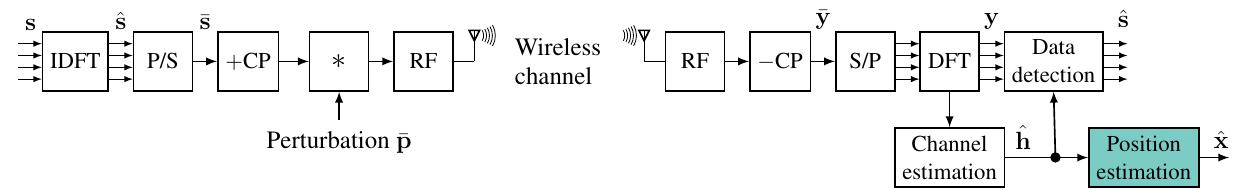}
    }
\vspace{0.3cm}
    \caption{OFDM communication system. A wireless transmitter (left) generates symbols $\bms$ in the frequency domain that are converted to the time domain followed by prepending a cyclic prefix (CP). The signal is then transmitted over the wireless channel. After removing the CP and conversion into the frequency-domain, the receiver (right) generates an estimate $\hat\bmh$ of the wireless channel, which is used for data detection as well as wireless positioning. Our attack pre-convolves the time-domain signal with a perturbation sequence $\bar\bmp$ that is kept constant for the duration of the entire OFDM packet.
    }
    \label{fig:ofdmsystem}
\end{figure*}

\subsection{System Model}\label{sec:communication system}
We consider an OFDM-based wireless communication system as illustrated in  \fref{fig:ofdmsystem} in which a single-antenna mobile user transmits information to an infrastructure BS or AP. The case with multiple receive antennas is discussed below.\footnote{Extending this model to the multiple-input multiple-output (MIMO) case is straightforward; see, e.g.,~\cite{gonultacs2021csi}.}
The model described in the following ignores, besides noise, any system and hardware impairments that are present in real-world transceivers. Nonetheless, our experiments in \fref{sec:results} with real-world CSI measurements include the effect of such impairments. 

We assume that a mobile user transmits $W$ frequency-domain symbols contained in the vector $\bms \in \complexset^W$. 
The frequency-domain symbols are first converted into the time domain $\bar{\bms} \in \complexset^W$ using an inverse discrete Fourier transform (IDFT) according to $\bar\bms =\bF^H\bms$, where $\bF\in\complexset^{W\times W}$ is the (unitary) DFT matrix. Then, a so-called cyclic prefix (CP) of length $C$ is prepended; the CP repeats the last $C$ entries of $\bar{\bms}$ to obtain $C+W$ time-domain samples that are then transmitted sequentially over the wireless channel. 

After time synchronization, the receiver first removes the cyclic prefix in the received time-domain samples.
The wireless channel can then be modeled with the following baseband input-output relation between the transmitting user and receiver:
\begin{align} \label{eq:inputoutputrelation}
\bar{\bmy}  = \bar{\bmh} \circledast  \bar{\bms} + \bar{\bmn}.
\end{align}  
Here, $\bar{\bmy}\in\complexset^{W}$ is the time-domain receive vector, $\bar{\bmh}\in\complexset^L$ is the channel's impulse response (consisting of $L\leq C+1$ nonzero taps), and $\bar{\bmn}\in\complexset^{W}$ models additive noise. 
The receiver then applies a $W$-point DFT to $\bar{\bmy}$ resulting in the frequency-domain receive vector $\bmy\in\complexset^W$, which leads to the following input-output relation of the wireless channel in the frequency domain~\cite{li2006orthogonal}:
\begin{align} \label{eq:yhs}
\bmy=\bmh \odot \bms + \bmn.
\end{align}
Here, $\bmh$ is the channel's transfer function, given by the DFT of the zero-padded impulse response~$\bar{\bmh}$, and $\bmn\in\complexset^W$ is additive noise. 
We see from \fref{eq:yhs} that OFDM converts the wireless channel into $W$ orthogonal subchannels, which significantly reduces the complexity of detecting the transmitted data in the presence of inter-symbol interference. 

Prior to data detection, the receiver needs to estimate the transfer function $\bmh$; this is commonly accomplished by transmitting a pre-defined pilot sequence, e.g., $\bmt\in \{-1,+1\}^W$, that is known to the receiver. By computing  $\hat{\bmh} = \bmy\odot\bmt$, the receiver can estimate the transfer function and use it to detect the transmitted data. 
We note that in the presence of $B$ antennas at the BS or AP, one estimates $B$ channel vectors $\hat{\bmh}_b$, $b=1,\ldots,B$, one for every receive antenna. The collection of the $B$ channel vectors results in an estimated $B\times W$ CSI matrix $\hat\bH=[\bmh_1,\dots,\bmh_{B}]^T$, which can be used for wireless positioning. 

\subsection{CSI-Based Positioning using Neural Networks}
\label{sec:positioningpipeline}

CSI-based positioning using DNNs first converts the estimated CSI matrix $\bH$ into a CSI feature vector $\bmf\in\reals^{d}$ using a feature-extraction function $\mathsf{F}:\complexset^{B\times W} \to \reals^d$; here, $d$ is the feature dimension.
CSI features render the positioning system resilient to system and hardware impairments, such as time-synchronization errors or carrier frequency offset~\cite{huawei2020paper,gonultacs2021csi}.
The CSI feature~$\bmf$ is then fed into a positioning DNN $\mathsf{P}:\reals^d\to \reals^D$ that produces a location estimate $\hat{\bmx}\in\reals^D$ of the transmitting device at location $\bmx\in\reals^D$; here, $D$ is the spatial dimension, which is commonly~$2$ or~$3$. 
The positioning DNN is typically trained off-line from a large database containing CSI feature and ground-truth location estimate pairs, which can be acquired using a reference positioning system. 

{In what follows, we consider two distinct CSI feature-extraction functions in order to study the trade-off between positioning accuracy and resilience to (adversarial) perturbations.}
The first function~$\mathsf{F}_1$ was developed in~\cite{huawei2020paper} and~\cite{gonultacs2021csi} to achieve state-of-the-art CSI-based positioning accuracy in indoor and outdoor scenarios, respectively. 
The second function $\mathsf{F}_2$ is simpler and typically achieves lower positioning accuracy, but will, as shown in  \fref{sec:results}, improve robustness against (adversarial) attacks in some scenarios. 
In the ensuing discussion, we will often use $\mathsf{F}$ to refer to either of the two CSI feature extraction functions $\mathsf{F}_1$ or $\mathsf{F}_2$. 
The details of these two CSI feature extraction functions are as follows:

\subsubsection*{Feature 1}
The feature-extraction pipelines from \cite{huawei2020paper,gonultacs2021csi} first take the inverse DFT of the rows of the CSI matrix according to $\bar{\bH}= \bH\bF^H$ in order to convert the $B$ transfer functions into the delay domain. One then computes a 2-dimensional (instantaneous) autocorrelation to obtain $\bR=\bar{\bH} \ast \bar{\bH}^*$ which doubles the number of rows and columns. Finally, the feature-extraction pipeline vectorizes the resulting autocorrelation $\bmr=\mathrm{vec}(\bR)$, extracts the real and imaginary parts $\bmr_R = [  \Re\{\bmr\}^T , \Im\{\bmr\}^T ]^T$, and obtains the unit-length CSI feature vector $\bmf_1 = \bmr_R/\|\bmr_R\|$.
%

\subsubsection*{Feature 2}
The second feature-extraction pipeline first takes the inverse DFT of the rows of the CSI matrix $\bar{\bH}= \bH\bF^H$. We then vectorize  the matrix and take the entry-wise absolute values $\bmh_v = |\mathrm{vec}(\bar{\bH})|$. The unit-length CSI feature vector corresponds to $\bmf_2 = \bmh_v/\|\bmh_v\|$.

In order to map CSI features to position estimates, we use a DNN that generates a so-called probability map $\bmm\in [0,1]^K$, where $K$ is the number of pre-defined grid points $\bmg_k\in\reals^D$, $k=1,\ldots,K$, covering the area of interest~\cite{gonultacs2021csi}. Each entry~$m_k$ indicates the likelihood of the user being at the specific grid point $\bmg_k\in\reals^D$ in  space. By computing $\hat\bmx = \sum_{k=1}^K \bmg_k m_k$, we obtain the final location estimate. 
We note that the convex hull spanned by the grid points $\{\bmg_k\}_{k=1}^K$ bounds the possible position estimates. Consequently, this approach limits the maximum perturbation of any (adversarial) attack on such probability-map-based positioning pipelines.
For positioning, we use neural networks with five dense layers with batch normalization (BN) in the first two layers. The first four layers use rectified linear units (ReLUs); the last layer uses softmax activations which generate the probability map~$\bmm$ for the $K$ predefined grid points. The number of activations per layer depends on the dataset and is shown in Figs.~\ref{fig:modeloutdoor} and \ref{fig:modelindoor}.
We use a binary cross entropy loss in order to train the network~$\mathsf{P}$ with reference probability maps generated from ground-truth position estimates.  
All positioning networks, adversarial examples, and defensive mechanisms were trained on a 48\,GB RAM Intel Core i7-7700K PC with an NVIDIA Quadro P6000 GPU.

\section{Adversarial Attacks and Defensive Mechanisms}
{We now detail our attack mechanism and propose adversarial as well as random attacks with varying levels of knowledge on the CSI-based positioning pipeline. We also propose defensive mechanisms that improve robustness of off-device CSI-based positioning pipelines against the proposed attacks.}

\subsection{Transmit-Side Perturbation}\label{sec:transsidepert}
Since one cannot simply perturb the measured CSI at the BS or AP side, a practical defense must inevitably originate at the transmitter side. 
As depicted in \fref{fig:ofdmsystem}, our approach pre-convolves the time-domain transmit signal with a perturbation sequence $\bar\bmp\in\complexset^{L_p}$ of length $L_p$ that is held constant for the entire duration of one OFDM packet transmission (which includes OFDM symbols for time and frequency synchronization, channel estimation, and payload data).
This approach causes the receiver to only observe the joint effect of the perturbation sequence and the channel's impulse response, i.e., is only able to estimate the combined impulse response $\bar\bmc = \bar\bmh\ast\bar\bmp$. Hence, the transmitter is able to efficiently alter the measured CSI, which is what we use as our defense against CSI-based positioning systems. 
As long as the combined impulse response duration satisfies $L+L_p\leq C+1$, orthogonality among the subcarriers is guaranteed and hence, our approach remains to be standard compliant and reliable transmission is generally possible if we enforce a unit-norm power constraint on the perturbation sequences.
What is more, our results in \fref{sec:results} reveal that short perturbation sequences (e.g., $L_p=16$ or much less) are already sufficient in order to significantly degrade the positioning accuracy.

In order to understand the impact our approach has on the reliability, we study its effect on the frequency-domain input-output relation in \fref{eq:yhs}, which, after perturbation, is given by 
\begin{align} \label{eq:yhs_perturbed}
\bmy=(\bmh \odot \bmp) \odot \bms + \bmn = \bmc \odot \bms +\bmn.
\end{align}
Here, $\bmp\in\complexset^W$ is the transfer function of the perturbation sequence given by the DFT of a zero-padded version of $\bar{\bma}$ according to $\bmp=\sqrt{W}\bF[\bar\bmp^T,\bZero_{(W-L_p)\times1}^T]^T$, and $\bmc=\bmh \odot \bmp$ models the combined effect of the perturbation sequence and the wireless channel.
Note that our attack is equivalent to pre-multiplying the frequency-domain symbols $\bms$ with the transfer function $\bmp$.\footnote{{A na\"ive perturbation of the transmit signal in the frequency domain, i.e., by directly manipulating $\bmp$ as done in, e.g., \cite{abanto2020wifi,COMINELLI2021107970,zhou2019adversarial}, may result in  perturbation sequences of length $L_p=W$ that violate orthogonality among subcarriers (due to the finite cyclic prefix length $C$) and, therefore, may cause serious performance degradation.} Our proposed time-domain approach ensures that the combined impulse response length is no longer than $L+L_p$.}

{In order to assess the impact of the proposed transmit-side perturbation on transmission reliability, without having to chose a specific modulation and coding scheme, we consider the OFDM channel capacity. The average per-subcarrier rate is given by~\cite{li2006orthogonal}:}
\begin{align} \label{eq:rate}
R \define \frac{1}{|\Omega|} \sum_{w\in\Omega} \log_2\!\left(1+\frac{|h_w p_w|^2 E_s}{\No}\right)\!.
\end{align}
Here, $\Omega\subseteq\{1,\ldots,W\}$ is the set of subcarriers reserved for data transmission, $E_s$ is the symbol power, and $\No$ the thermal noise power at the receiver. 
By manipulating the time-domain signals with the perturbation sequence $\bar\bmp$, one alters the combined frequency-domain gain $|h_w p_w|^2$ per subchannel, which will affect the rate $R$. First, we have to ensure that the transmit power remains unaffected by our perturbation attack. As mentioned above, this  can be accomplished by enforcing $\|\bmp\|=1$. Second, we would like that \fref{eq:rate} is as large as possible; this  can be accomplished by including a rate-regularization term when computing adversarial perturbation sequences. 

In the case of $B$ receive antennas, the proposed perturbation approach affects each row of the estimated CSI matrix in the following way $\hat{\bH}[\bmp]=[\bmh_1 \odot \bmp ,\dots,\bmh_{B}\odot \bmp]^T$, as every path between the transmitter and the $b$th receive antenna is perturbed in the same manner. 
The per-subcarrier rate $R$ in this scenario is obtained by replacing $|h_w p_w|^2$ in \fref{eq:rate} by $|p_w|^2\sum_{b=1}^B|H_{b,w}|^2$.

\subsection{On-Device Adversarial Attacks}\label{sec:extraction}
The remaining piece in our defense against CSI-based positioning is the generation of effective adversarial perturbation sequences, which we detail next.

\subsubsection*{White-Box Attack}
{As a gold standard, we first consider a \emph{white-box attack} in which the positioning DNN~$\mathsf{P}$ and the feature extraction function~$\mathsf{F}$ are both known to the adversary (the user), and one has access to the estimated CSI matrix~$\hat\bH$.} In this case, the location estimate is generated at the receiver by cascading both functions $\hat\bmx[\bmp]=\mathsf{P}(\mathsf{F}(\hat{\bH}[\bmp]))$, which depends on the time-domain perturbation $\bar\bmp$ via $\bmp=\sqrt{W}\bF[\bar\bmp^T,\bZero_{(W-L_p)\times1}^T]^T$. 
{One can now extract an adversarial perturbation sequence by approximately solving the optimization problem}
\begin{align} \label{eq:adversarial}
\bar\bmp_\text{adv} = \, & \argmax_{\tilde\bmp\in\complexset^{L_p}} \big\|\hat\bmx- \underbrace{\textsf{P}(\textsf{F}(\hat{\bH}[\sqrt{W}\bF[\tilde\bmp^T,\bZero_{(W-L_p)\times1}^T]^T]))}_{\hat\bmx_\text{adv}}\!\big\|^2 \notag \\ 
& \text{subject to} \,\, \|\tilde\bmp\|=1
\end{align}
{using projected gradient ascent, which aims at pushing the perturbed position estimate $\hat\bmx_\text{adv}$ as far away as possible from the unperturbed position estimate $\hat\bmx=\mathsf{P}(\mathsf{F}(\hat{\bH}))$.}
{We include a normalization constraint on the perturbation sequence $\tilde\bmp$, which ensures that the transmit power remains unaffected (see also our discussion in \fref{sec:transsidepert}).}
{Furthermore, since~\fref{eq:adversarial} only focuses on maximizing the positioning error, the per-subcarrier rate $R$ may deteriorate.} Hence, we add a term $\lambda R$ from \fref{eq:rate} to the objective in~\fref{eq:adversarial}, where the parameter $\lambda$ provides a trade-off between position perturbation and rate. 
{We note that due to the nonconvexity of problem \fref{eq:adversarial}, we resort to projected gradient ascent in order to find suitable perturbation sequences---optimality of the obtained sequences can, however, not be guaranteed. Therefore, to evaluate the efficacy of the  proposed attacks, we resort to simulations with real-world datasets in \fref{sec:results}.}

\subsubsection*{Transfer Attack}
{While the above white-box attack has the strongest impact on the positioning error (cf.~\fref{sec:results}), it also requires knowledge of the CSI feature extraction function $\mathsf{F}$, the positioning function $\mathsf{P}$, and the estimated CSI $\hat\bH$ which may be unavailable to the transmitting user.}
We therefore propose an alternative approach we call \emph{transfer attack}. This attack type assumes that the transmitter has access to an alternative positioning network~$\mathsf{P}_\text{alt}$ that has been learned with a different BS or AP, or was obtained from a simulated scenario. One then simply replaces the true model $\mathsf{P}$ in the adversarial attack~\fref{eq:adversarial} by~$\mathsf{P}_\text{alt}$ in order to extract the perturbation sequence from estimated CSI. 

\subsubsection*{Pool Attack}
The transfer attack still requires access to estimated CSI, which can, in certain systems, be extracted through channel reciprocity in a time-division duplexing system. In order to avoid knowledge of the estimated CSI altogether, we propose another approach called \emph{pool attack}.
This attack precomputes a pool of $T$ different perturbation sequences $\{\bar\bmp_t\}_{t=1}^T$ for randomly selected locations in the given area and then, selects {perturbation sequences from this pool uniformly and at random.}

\subsubsection*{Random Attack}
{We also consider a simple approach dubbed \emph{random attack}, which is no longer deterministic (as are the white-box and transfer attacks) but has the advantage of no longer requiring knowledge of the positioning function, CSI feature extraction function, or CSI estimates.}
Here, we randomly generate the entries of the $L_p$-length perturbation sequence $\bar\bmp$ by sampling the amplitudes and phases from i.i.d.\ uniform distributions in $[0,1]$ and  $[0,2\pi)$, respectively. {Concretely,  $\bar{p}_k=A_ke^{j\phi_k}$ with $A_k\sim\text{U}([0,1])$ and $\phi_k\sim\text{U}([0,2\pi))$ for $k=1,\ldots,L_p$, where $\text{U}$ denotes the uniform distribution.}
We then normalize the perturbation vector so that $\|\bar\bmp\|=1$.

\subsection{Improving Positioning Robustness using Adversarial \mbox{Training}} 
\label{sec:adversarialtraining}

Since we are proposing new methods to degrade the accuracy of off-device CSI-based positioning systems, it is natural to ask what will happen if the positioning  system is aware of such attacks and deploys suitable countermeasures.
To this end, we also investigate adversarial training techniques, which generally improve  model robustness by training the neural network parameters on particularly ``difficult'' examples~\cite{szegedy2014intriguing,shafahi2019adversarial}.
Although adversarial training typically incurs an accuracy loss for the inference of unperturbed data~\cite{engstrom2018evaluating}, it effectively improves the resilience of neural networks against challenging (or adversarially generated) input data.

In what follows, we consider a simple yet effective defense mechanism, where we, instead of training the positioning network with adversarial examples, we utilize training samples generated from the random attack as introduced in \fref{sec:extraction}.
{Concretely, during training of the positioning network $\textsf{P}$, for every pair of ground-truth probability map and output of the positioning network in each batch, we perturb the estimated CSI features as  $\textsf{F}(\hat{\bH}[\sqrt{W}\bF[\tilde\bmp^T,\bZero_{(W-L_p)\times1}^T]^T])$ as in the objective function of~\fref{eq:adversarial}.
Here, the transmit-side perturbation sequences~$\tilde\bmp$ are chosen at random using the statistical model used for the random attack proposed in \fref{sec:extraction}.}
The advantages of this approach are (i) simplicity and (ii) universality, as the off-device positioning system can be trained from randomly perturbed data only and the method is independent of the attack type deployed at the user side. 
An investigation of other, more sophisticated defensive mechanisms that leverage adversarial examples during training is left for future work. 

\section{Results}
\label{sec:results}

{Since the CSI feature extraction function $\mathsf{F}$ and the positioning function $\mathsf{P}$ are nonlinear, and the optimization problem in \fref{eq:adversarial} is nonconvex, a theoretical analysis of our attack and defense mechanisms is challenging. Therefore, we resort to numerical experiments with real-world measurement campaigns in a large outdoor 5G cellular scenario and an indoor IEEE 802.11ac WLAN scenario to investigate the cause-and-effect relationships of the proposed methods.}

\subsection{Outdoor 5G  Cellular Scenario}
\label{sec:outdoordataset}
\subsubsection*{Dataset, Models, and Training}

\begin{figure}
     \centering
\adjustbox{max width=\columnwidth,valign=t}{%
\includegraphics{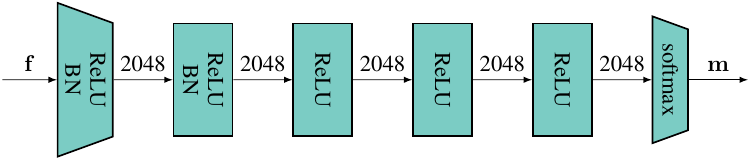}
    }
\vspace{0.05cm}
    \caption{
    Neural network architecture for the outdoor 5G cellular scenario. The input is a CSI feature $\bmf$ and the output is a $30\times 30$ probability map $\bmm$.
    }
    \label{fig:modeloutdoor}
\end{figure}

\begin{figure*}[tp]
    \centering
    \subfloat[Test-set GNSS locations \label{fig:scen_1}]
    {
    \adjustbox{max width=0.8\columnwidth,valign=t}{%
         \includegraphics{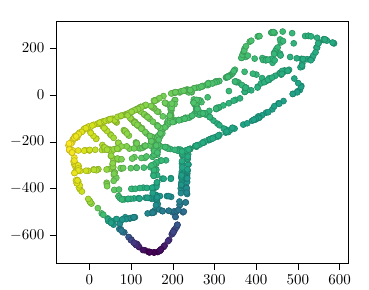}
         }
    }
    \hspace{1.2cm}
    \subfloat[Estimated locations w/o perturbation\label{fig:scen_2}]
    {
    \adjustbox{max width=0.8\columnwidth,valign=t}{%
         \includegraphics{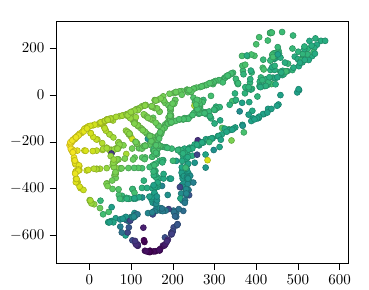}
         }
    }
    
    \subfloat[White-box attack \label{fig:scen_3}]
    {
    \adjustbox{max width=0.8\columnwidth,valign=t}{%
         \includegraphics{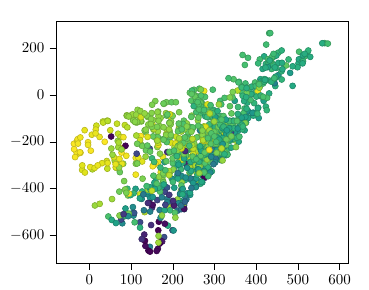}
         }
    }
     \hspace{1.2cm}
    \subfloat[Random attack \label{fig:scen_4}]
    {
    \adjustbox{max width=0.8\columnwidth,valign=t}{%
         \includegraphics{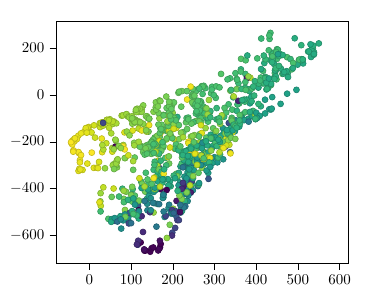}
         }
    }    
\vspace{0.1cm}
    \caption{Real locations, estimated locations without perturbations, as well as white-box and randomly perturbed locations with $L_p=16$  for the test-set of the outdoor 5G cellular dataset. The x and y axes are in meters and show the relative position to the BS which is located at $(x,y)=(0,0)$. The color gradients help to visualize the position perturbation.}
    \label{fig:Nokia_locations}
    \vspace{-0.2cm}
\end{figure*}

We start by evaluating the efficacy of our attacks with the measured outdoor dataset from the IEEE Communication Theory Workshop (CTW) 2020 data competition~\cite{arnold2019novel}, consisting of $N = 22\,128$ CSI estimates and GPS locations as ground-truth position estimates. 
The dataset was acquired in a large suburban area and a random subset of the test-set locations is shown in~\fref{fig:scen_1}. 
A mobile transmitter in the measurement area communicates with a $8\times8$ uniform rectangular array (URA) placed on a high building complex at a carrier frequency of $1.27$\,GHz and a bandwidth of $20$\,MHz.
Each estimated CSI matrix consists of $B=64$ antennas and $W =|\Omega|= 924$ active subcarriers; the CSI feature dimension is $d=117\,602$.
The first four layers of the positioning DNN have $2048$ activations and the output produces a $K=30\times30$  probability map, which also enforces a bounding box around the measurement area.
\fref{fig:modeloutdoor} illustrates the used neural network architecture. The numbers next to the arrows correspond to the number of activations.
In order to train the alternative positioning network $\mathsf{P}_\text{alt}$ for the transfer attack, we use the QuaDRiGa channel model~\cite{quadriga} in an urban micro scenario with the same system parameters (bandwidth, carrier frequency, BS location, etc.) as for the CTW dataset.

\begin{figure*}[tp]
    \centering
    \subfloat
    {
    \adjustbox{max width=0.8\columnwidth,valign=b}{%
        \includegraphics{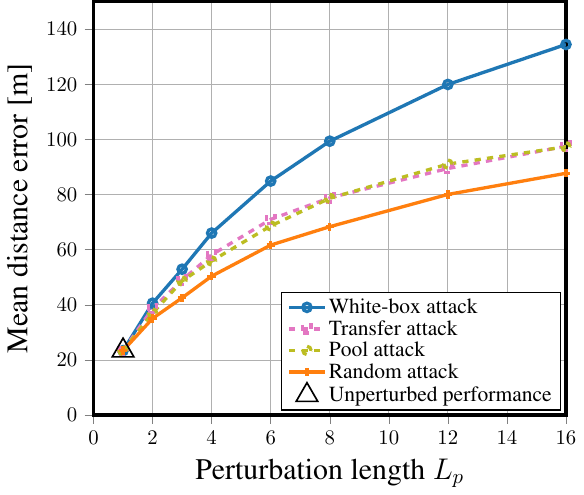}
        }
    }
    \hspace{1.2cm}
        \subfloat
    {
    \adjustbox{max width=0.8\columnwidth,valign=b}{%
        \includegraphics{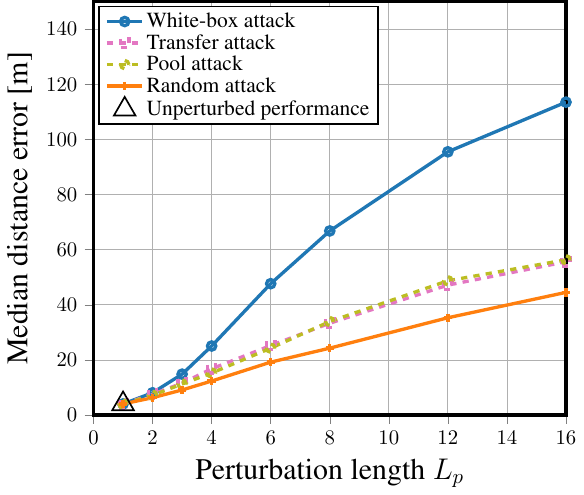}
        }
    }   
    \caption{Outdoor 5G  cellular dataset with Feature~1: mean distance error (left) and median distance error (right) on the test-set for different adversarial attacks and with varying perturbation length $L_p$.}
    \label{fig:Nokia_dataset}
\end{figure*}

\begin{figure*}[tp]
    \centering
    \subfloat
    {
    \adjustbox{max width=0.8\columnwidth,valign=b}{%
        \includegraphics{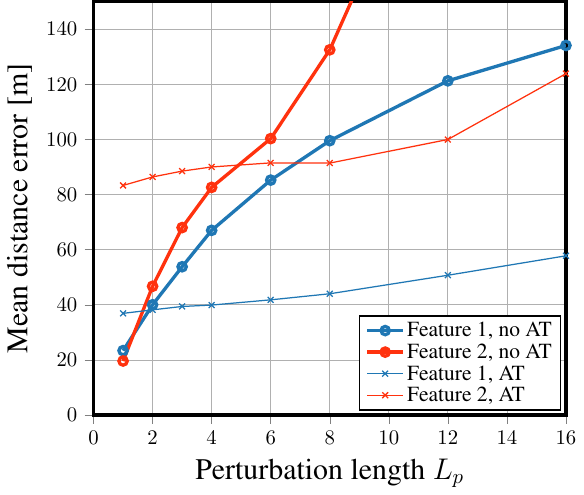}
        }
    }
    \hspace{1.2cm}
        \subfloat
    {
    \adjustbox{max width=0.8\columnwidth,valign=b}{%
        \includegraphics{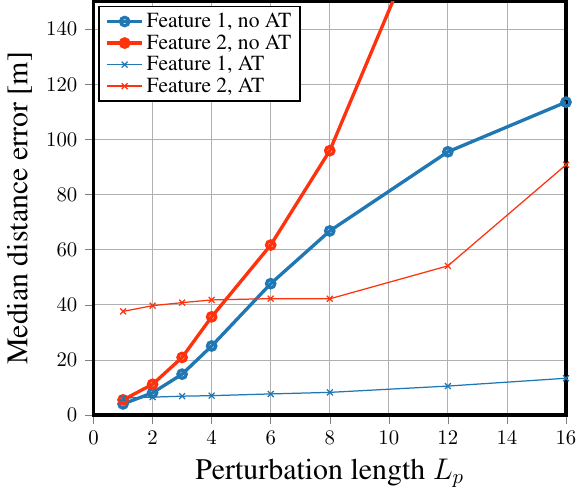}
        }
    }
    \caption{Comparison of the test-set mean and median distance errors after performing the white-box attack on a model trained with the 5G cellular dataset on both feature functions $\mathsf{F_1}$ and $\mathsf{F}_2$ with and without adversarial training (AT). }
    \label{fig:Nokia_dataset_alt}
\end{figure*}

\subsubsection*{Results}
\fref{fig:Nokia_dataset} shows the mean and median distance errors in meters, depending on the perturbation length $L_p$ with CSI Feature~1. 
For $L_p=1$, the unperturbed and perturbed performance is the same (approximately 23\,m mean and 4\,m median distance errors) as the  feature extraction function~$\mathsf{F}_1$ in \fref{sec:positioningpipeline} ignores global phase and amplitude changes. 
By increasing the perturbation length, the mean and median distance errors increase rapidly, but the probability map bounds the maximum distance error to approximately $600$\,m, resulting in a saturation behavior as $L_p$ approaches $16$. 
As expected, the white-box attack causes the largest mean and median distance errors, {as it has perfect knowledge of the positioning DNN~$\mathsf{P}$, the CSI feature extraction function $\mathsf{F}$, and the estimated CSI matrix $\hat\bH$, which enables this attack to perfectly craft adversarial perturbation sequences. In contrast, the random attack has the least impact but requires no knowledge of the positioning DNN, the CSI feature extraction function, and estimated CSI. Nonetheless, the random attack still induces a  remarkable 68\,m average and 24\,m median distance error for a short perturbation length of $L_p=8$.}
\fref{fig:scen_3} and \fref{fig:scen_4} illustrate the effect of these two perturbations on a subset of the location estimates with $L_p=16$. While both methods are effective in improving location privacy, the color gradients indicate that coarse localization remains to be possible; see \fref{sec:limitations} for a detailed discussion on the limitations of our work. 
The transfer and pool attacks provide improvements over the random attack, albeit only marginally. This indicates that having accurate knowledge of the model and CSI available to the white-box attack would be advantageous, but the other attacks can cause distance errors in excess of  $90$\,m, which enable substantial location privacy improvements. 
Note that for this dataset, we do not have accurate SNR estimates, which makes it difficult to study the per-user rates $R$. We shed light on this aspect for the indoor dataset discussed in \fref{sec:indoorwlanscenario}. 

\fref{fig:Nokia_dataset_alt} demonstrates the effect of adversarial training as proposed in \fref{sec:adversarialtraining} to improve resilience of the positioning network to the proposed white-box attack. 
We compare Feature~1 with Feature 2 while learning the parameters of the positioning network with and without adversarial training (AT).
As expected, AT renders the positioning network more resilient to our attacks for increasing perturbation lengths $L_p$. 
However, improved resilience comes a the cost of sacrificing positioning accuracy in absence of any attack as it can be seen for $L_p=1$.
We furthermore see that Feature~1, which was also used in \fref{fig:Nokia_dataset}, is more robust to adversarial perturbations, whereas Feature~2, which performs (slightly) better for unperturbed data ($L_p=1$), suffers significantly in the presence of the white-box attack. 
{Evidently, the CSI feature extraction function has a significant impact on the trade-off between attack robustness and positioning accuracy, and we will show next how the impact can differ in an indoor scenario.}

\subsection{Indoor WLAN Scenario}
\label{sec:indoorwlanscenario}
\subsubsection*{Dataset, Models, and Training}

\begin{figure}[tp]
     \centering
     \adjustbox{max width=\columnwidth,valign=t}{%
\includegraphics{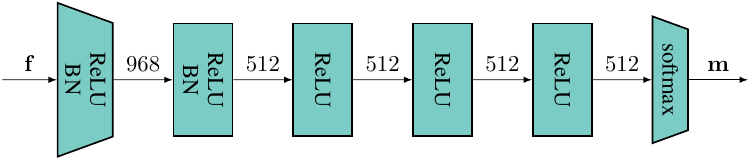}
    }
\vspace{0.05cm}
    \caption{
    Neural network architecture for the indoor WLAN scenario scenario. The input is a CSI feature $\bmf$ and the output is a $22\times 22$ probability map $\bmm$.
    }
    \label{fig:modelindoor}
\end{figure}

We now evaluate our attacks on the measured indoor WLAN dataset from~\cite{gonultacs2021csi} consisting of $N = 50\,000$ CSI estimates and ground-truth locations obtained using a VICON precision positioning system \cite{vicon}.
The dataset was acquired using a robot moving in an area of $4.2\times 2.9$ m$^2$ of a lab; due to the small area, only small perturbations are to be expected.
A mobile transmitter communicates with two $4$-antenna IEEE 802.11ac WLAN APs (AP1 and AP2) both operating at a carrier frequency of $5$\,GHz and a bandwidth of $80$\,MHz under line-of-sight conditions. 
The positioning network~$\mathsf{P}$ was trained for AP1 and the alternative network~$\mathsf{P}_\text{alt}$ for the transfer attack was trained for AP2, which is at an opposite location of the laboratory environment.
The training and test sets were recorded separately. 
Each estimated CSI matrix consists of $B=4$ antennas and $W =|\Omega|= 234$ active subcarriers; the CSI feature dimension is $d=3\,570$. The SNR exceeded 10\,dB for all measurements.  
\fref{fig:modelindoor} illustrates the used neural network architecture.

\begin{figure*}[tp]
    \centering
    \subfloat 
    {
    \adjustbox{max width=0.8\columnwidth,valign=b}{%
        \includegraphics{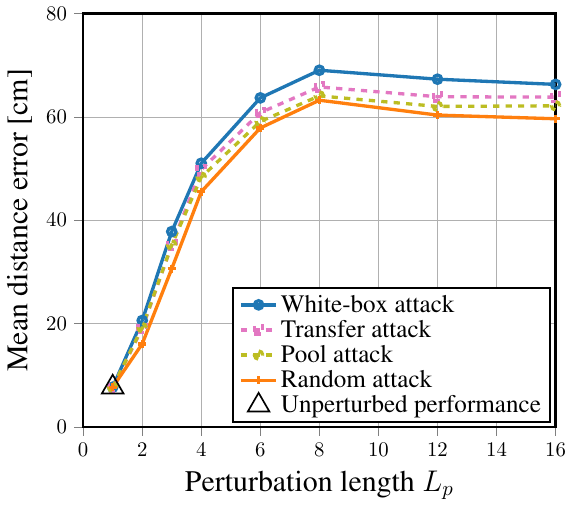}
        }
    }
    \hspace{1.2cm}
        \subfloat 
    {
    \adjustbox{max width=0.8\columnwidth,valign=b}{%
        \includegraphics{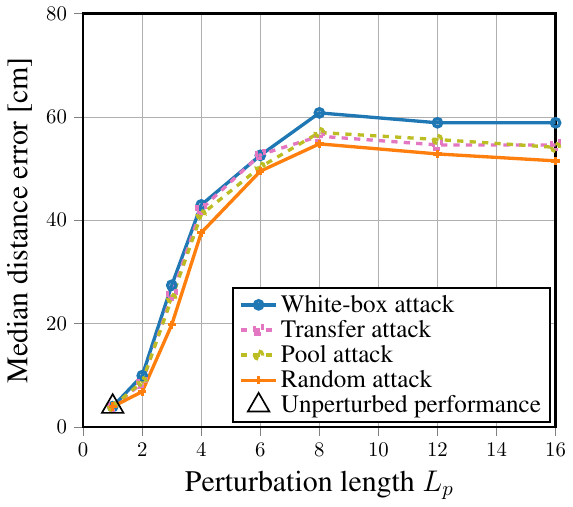}
        }
    }
    \caption{Indoor WLAN dataset: mean distance error (left) and median distance error (right) on the test-set for different adversarial attacks and with varying perturbation length $L_p$.}
    \label{fig:Indoor_l2d}
\end{figure*}

\begin{figure*}[tp]
    \centering
    \subfloat
    {
    \adjustbox{max width=0.8\columnwidth,valign=b}{%
        \includegraphics{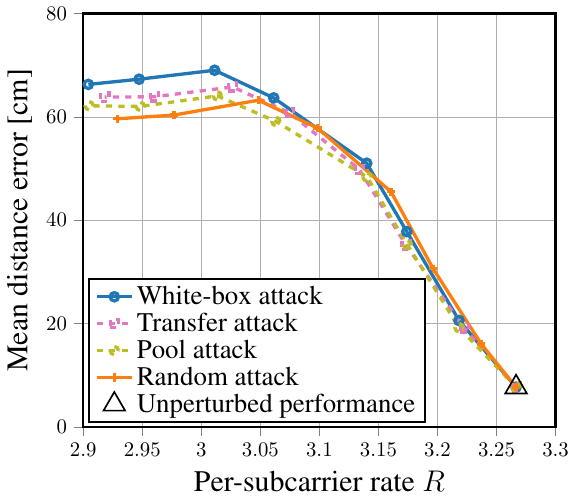}
        }
    }
    \hspace{1.2cm}
    \subfloat
    {
    \adjustbox{max width=0.8\columnwidth,valign=b}{%
        \includegraphics{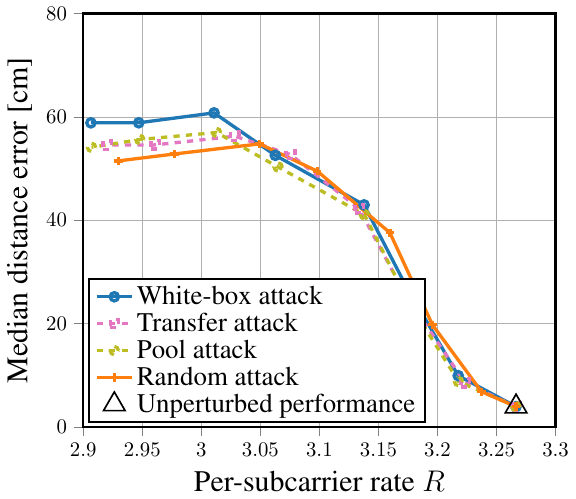}
        }
    }
    \caption{Indoor WLAN dataset: mean distance error (left) and median distance error (right) on the test-set for different adversarial attacks dependent on the per-subcarrier rate $R$.}
    \label{fig:Indoor_2rate}
\end{figure*}

\begin{figure*}[tp]
    \centering
    \subfloat 
    {
    \adjustbox{max width=0.8\columnwidth,valign=b}{%
        \includegraphics{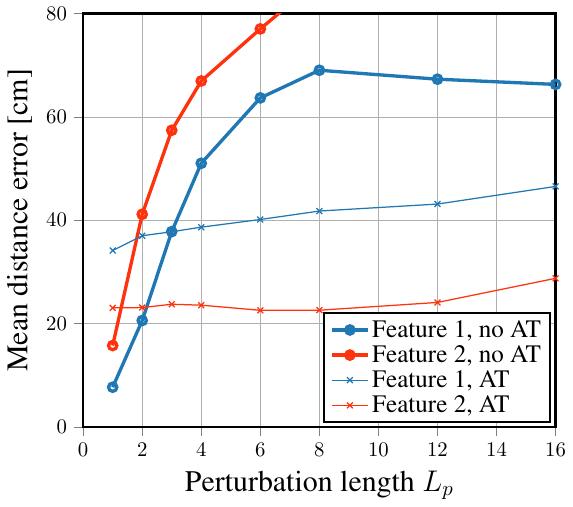}
        }
    }
    \hspace{1.2cm}
        \subfloat 
    {
    \adjustbox{max width=0.8\columnwidth,valign=b}{%
        \includegraphics{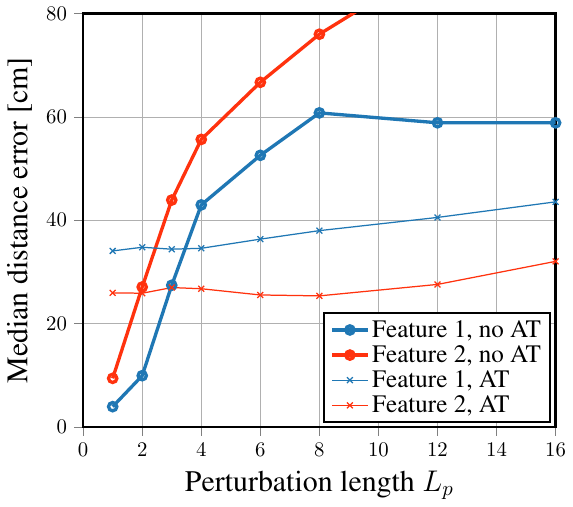}
        }
    }
    \caption{Comparison of the test-set mean and median distance errors after performing the white-box attack on a model trained with the indoor dataset on both feature functions $\mathsf{F_1}$ and $\mathsf{F}_2$ with and without adversarial training (AT). }    
    \label{fig:exIndoor_l2d_alt}
\end{figure*}

\subsubsection*{Results}
\fref{fig:Indoor_l2d} shows the mean and median distance errors in centimeters, for varying perturbation lengths~$L_p$ with Feature~1. For $L_p=1$, the unperturbed and perturbed performance is the same (approximately 8\,cm mean and 4\,cm median distance error). 
As for the indoor dataset, increasing the perturbation length causes the mean and median distance errors to increase rapidly but saturate already for $L_P\geq8$. The saturation effect comes from the probability maps which bound the possible position estimates in the relatively small area. 
As for the outdoor dataset, the white-box attack causes the strongest perturbation {as it has perfect knowledge of the positioning pipeline,} whereas the random attack has the lowest impact {but without requiring any model knowledge}. The random attack can still achieve distance errors of more than 63\,cm on average and 54\,cm on median. The white-box attack has a 5\,cm to 6\,cm advantage over the random attack; the other two attacks yield 2\,cm to 3\,cm.
{Note that the impacts of adversarial attacks to the indoor scenario are significantly smaller compared to the outdoor scenario in \fref{sec:outdoordataset}, which is due to the facts that (i) the probability maps used in our positioning pipeline bound the space of position estimates (and the space is much smaller for the indoor dataset) and (ii) the specific scenario, which determines the CSI dataset, affects the efficacy of our attacks.}

\fref{fig:Indoor_2rate} shows the impact of our attacks to the per-subcarrier rate $R$. We see that the unperturbed model for $L_p=1$ achieves a rate of approximately $R=3.27$\,b/subcarrier at 10\,dB SNR. By increasing the distance errors with a perturbation length of $L_p=8$, the rate drops only marginally, i.e., to about $R=3$\,b/subcarrier, but causes positioning errors of over $50$\,cm.
From this result we infer that the impact of adversarial attacks on quality-of-service is negligible and unlikely to be noticed, especially since automatic retransmission often hides such marginal performance losses.

\fref{fig:exIndoor_l2d_alt} demonstrates the effect of adversarial training as proposed in \fref{sec:adversarialtraining} to improve resilience against perturbation attacks. 
Again, we compare Feature 1 with Feature~2 while training the positioning network with and without adversarial attacks in mind.
As for the outdoor dataset, we see that adversarial training renders the positioning network more resilient to attacks for growing perturbation lengths $L_p$, but improved resilience comes a the cost of sacrificing positioning accuracy in absence of any attack.
Here, we also see that Feature 1 yields improved positioning accuracy compared to Feature 2 in absence of any attack, but Feature 2 turns out to be more robust than Feature 1 with adversarial training, which is in contrast to what we observed in the outdoor scenario. {These observations again demonstrate the presence of a trade-off between positioning accuracy and robustness to adversarial attacks, and highlights the importance of the chosen CSI features.}

\section{Limitations and Implications}
\label{sec:limitations}

We now discuss the limitations and implications of our attacks and defenses. 

\subsection{Limitations of Our Work}

While our work proposes novel means to attack CSI-based off-device positioning systems and suitable countermeasures, the proposed methods  suffer from several limitations. 
One key drawback is that the proposed white-box attack requires knowledge of the {entire positioning pipeline} and the CSI, which is difficult to acquire in practice---our other attacks require less or no knowledge, but also are less effective. 
Hence, there is room for more effective attacks that require less knowledge but achieve similar location privacy as the proposed reference white-box attack. 
{Furthermore, due to the nonconvexity of the optimization problem in \fref{eq:adversarial}, our approach to extract perturbation sequences for the white-box attack with projected gradient ascent does \emph{not} guarantee optimality---alternative optimization algorithms may produce better results.}
{Also, we have not investigated approaches that \emph{learn} CSI feature extraction functions with the purpose of providing increased robustness to adversarial positioning attacks---such approaches, e.g., based on the CSI feature learning approach put forward in \cite{gonultacs2021feature}, are left for future work.}

Our results also indicate that one cannot prevent localization completely while enabling communication. 
In particular, \fref{fig:Nokia_locations} demonstrates that our attacks indeed prevent precise localization, but the perturbed locations with our most powerful white-box attack still capture positions on a larger-scale quite accurately. 
What is more: As soon as a device connects with a BS or AP, one will always know that the transmitter will be in its vicinity---our attacks cannot prevent that. 
The recent trend towards network densification (i.e., placing more BSs and APs in space) and communication at millimeter-wave (mmWave) frequencies further aggravate this issue.
Network densification will enable more accurate location estimates simply by processing connectivity information. Furthermore, mmWave communication typically operates under line-of-sight conditions, which further restricts the possible transmitter positions that are in range of a BS or AP. 
In addition, jointly processing CSI of one device at multiple BSs and APs to extract location estimates has been shown to significantly improve localization accuracy~\cite{gonultacs2021csi}. We speculate that such multi-point localization methods will be even harder to attack, as the same perturbation will be measured by multiple receivers at different locations in space. 
A possible defense might be to equip the transmitter with multiple antennas and to perform adversarial beamforming (e.g., by steering the signal into specific directions). Such  multi-antenna attacks might prove more effective than the ones proposed here and may also be able to fool positioning systems that process AoA and/or ToF, but may degrade the quality-of-service as one has to use sub-optimal beams. 
However, if one fuses multi-point CSI estimates with other sensor modalities, e.g., from cameras or radar systems, all such defensive mechanisms are likely to become ineffective, necessitating  fundamentally different attack strategies.

Furthermore, our randomized training method that renders off-device positioning systems more resilient to adversarial and random attacks indicate that one can (often significantly) improve robustness. While our approach has shown that this comes at the cost of reduced position accuracy,  {it} is possible that improved adversarial training techniques can be developed which may overcome this drawback. We leave a detailed investigation of this topic for future work.

\subsection{Future Implications}

Evidently, off-device wireless positioning threatens user privacy and security, and future communication systems (e.g., operating at mmWave frequencies) will further facilitate the extraction of accurate location estimates.
A pragmatic (albeit na\"ive) approach to prevent CSI-based positioning is to pass legislation that heavily regulates such off-device positioning technologies altogether. However, due to the facts that (i) CSI-based positioning is entirely passive and, hence, difficult (if not impossible) to detect at the transmitter side and (ii) it is notoriously difficult (and costly) to re-engineer a wireless communication receiver (e.g., a BS or AP) in order to determine whether it includes positioning capabilities or not, simply relying on laws or regulations might be insufficient. 
We believe that a reasonable path forward is to combine regulations with on-device perturbation attacks that can be turned on when~needed.

A more effective approach would be to consider location privacy already in the physical layer specification of future wireless systems. Possible techniques might include encrypted pilot sequences (known to the transmitter and receiver, but not to an eavesdropper), which is not part of most wireless standards. 
However, blind channel estimation~\cite{shin2007bilindchest} would still be able to extract accurate CSI if the transmitted packet and the channel's coherence time are sufficiently long.  
Hence, we believe that fundamentally new privacy mechanisms must be developed that prevent (or mitigate the efficacy of) off-device~localization.

\section{Conclusions}
We have shown that emerging off-device positioning systems that process measured CSI with neural networks cause serious privacy and security concerns. 
To address this issue, we have proposed simple yet effective attack mechanisms that perform an on-device convolution of the transmit signal with specifically-designed adversarial or random sequences.
Our results demonstrate that the positioning errors for real-world cellular outdoor and WLAN indoor systems can be increased substantially for short perturbation sequences while remaining standard compliant and causing a minimal loss in quality-of-service. 
Our results also show that adversarial training can mitigate the impact of our attacks at the cost of reduced accuracy for unperturbed transmissions. 

In summary, our results are encouraging, but reveal that the dichotomy between communication reliability and location privacy is hard to resolve as the acquisition of CSI is inherently linked to wireless transmission. 
We therefore expect that emerging communication systems operating at mmWave frequencies and with a denser deployment of BSs or APs that jointly process CSI for positioning, will render the design of effective adversarial attacks even more challenging. 
Hence, the design of novel attack and security mechanisms will be required, which ensure location privacy in future wireless systems that enable CSI-based off-device positioning.  

\balance

\bibliographystyle{IEEEtran}
\bibliography{IEEEabrv,VIPabbrv,jsacbib}

\end{document}